\documentclass[11pt]{article}
\usepackage{epsfig,bm,amsmath,amsfonts,algorithm,algpseudocode}
\title{Generalized $k$-Means in GLMs with Applications to the Outbreak of COVID-19 in the United States}
\author{Tonglin Zhang \thanks{Department of Statistics, Purdue University; 250 North University Street, West Lafayette, IN 47907-2066; email: tlzhang@purdue.edu} and Ge Lin\footnote{Department of Environmental and Occupational Health, University of Nevada Las Vegas, Las Vegas, NV 89154, Email: ge.kan@unlv.edu}}
\def\qed{\hfill$\diamondsuit$}

\newtheorem{thm}{Theorem}
\newtheorem{cor}{Corollary}
\newtheorem{lem}{Lemma}
\begin{document}
\maketitle
\def\eqalign#1{\null\,\vcenter{\openup\jot\ialign
              {\strut\hfil$\displaystyle{##}$&$\displaystyle{{}##}$
               \hfil\crcr#1\crcr}}\,}


\begin{abstract}

Generalized $k$-means can be incorporated with any similarity or dissimilarity measure for clustering. By choosing the dissimilarity measure as the well known likelihood ratio or $F$-statistic, this work proposes a method based on generalized $k$-means to group statistical models. Given the number of clusters $k$, the method is established under hypothesis tests between statistical models. If $k$ is unknown, then the method can be combined with GIC to automatically select the best $k$ for clustering. The article investigates both AIC and BIC as the special cases. Theoretical and simulation results show that the number of clusters can be identified by BIC but not AIC. The resulting method for GLMs is used to group the state-level time series patterns for the outbreak of COVID-19 in the United States. A further study shows that the statistical models between the clusters are significantly different from each other. This study confirms the result given by the proposed method based on generalized $k$-means. 
\end{abstract}


\section{Introduction}
\label{sec:introduction}

Generalized $k$-means, including both $k$-means and $k$-medians as special cases, can be incorporated with any similarity or dissimilarity measure to group objects (or observations). The similarity or dissimilarity measure can be very general. In this work, we specify the dissimilarity measure as the well known likelihood ratio or $F$-statistic, such that our method can be used to group statistical models. In particular, if each object contains a vector for the response and a design matrix for the explanatory variables, then statistical models can be used to describe the relationship between the response and explanatory variables within the object. A clustering problem arises if we want to group the statistical models between objects. This problem can be solved by generalized $k$-means. The current research proposes the  method and use it to group the state-level time series patterns for the outbreak of COVID-19 in the United States.

The outbreak of COVID-19 has become a worldwide pandemic since March 2020. According to the website of the World Health Organization (WHO), until July 31, the outbreak has affected over $200$ countries with almost eighteen million confirmed cases and seven hundred thousand deaths in the entire world. Among those, the United States has almost five million confirmed cases and one hundred sixty thousand deaths. To understand the outbreak in the United States, we compare the time series patterns for daily new cases in the fifty states and Washington DC. We find that some of the patterns are similar to each other and some of them are far away from each other, implying that we need to have a clustering method to group these patterns. As statistical models are involved, we cannot use traditional $k$-means. We recommend using generalized $k$-means.

Clustering is one of the most popular unsupervised statistical learning methods for unknown structures. Clustering methods are often carried out by a similarity or dissimilarity measure between objects such that they can be grouped into a few clusters. The purpose of clustering is to make objects within clusters mostly homogeneous and objects between clusters mostly heterogeneous. In the literature, one of the most well known clustering methods is the $k$-means. It assigns each object to the cluster with the nearest mean. Based on a given $k$, the $k$-means provides $k$ clusters according to $k$ centers. The $k$ centers are solved by minimizing the sum-of-squares (SSQ) criterion, which is derived based on the Euclidean distance between objects in the data. The SSQ criterion in the $k$-means can be replaced by any similarity or dissimilarity measure. A method called the generalized $k$-means is proposed~\cite{bock2008,soheily-khah2016}. Because the choice of the similarity or dissimilarity measure is flexible, generalized $k$-means can be extended to any divergence measure for clustering. 

Many clustering methods have been proposed in the literature. Examples include hierarchical clustering~\cite{zhao2005}, fuzzy clustering~\cite{trauwaert1991}, density-based clustering~\cite{kriegel2011}, model-based clustering, and partitioning clustering. Model-based clustering is usually carried out by EM algorithms or Bayesian methods under the framework of mixture models~\cite{fraley2002,lau2007}. Partitioning clustering can be interpreted by the centroidal Voronoi tessellation method in mathematics~\cite{du2002}. It can be further specified to $k$-means~\cite{forgy1965,hartigan1979,lloyd1982,macqueen1967}, $k$-medians~\cite{charikar2002}, and  $k$-modes~\cite{goyal2017}, where $k$-means is the most popular. To implement those, one needs to express observations of the data in a metric space, such that a distance measure can be defined. Several approaches have been developed to specify the distance measure. A review of these can be found in \cite{johnson2002}, P 670. 

\begin{figure}
\centerline{\rotatebox{270}{\psfig{figure=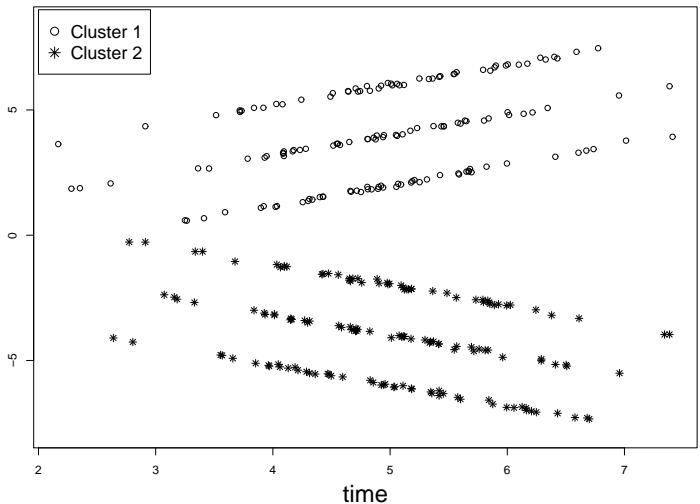,width=3.0in,}}}
\caption{\label{fig:six regression lines}Generalized $k$-means clustering for six regression lines.}
\end{figure}

Challenges appear when we want to group the time series patterns for the outbreak of COVID-19 in United States. Suppose that the time series patterns from individual states have been analyzed by statistical models. Then, we need to study the relationship between these models. It is inappropriate to directly compare their coefficients because of disparity. To overcome the difficulty, we recommending using likelihood ratio statistics or an $F$-statistics derived based on hypothesis tests for the relationship between these models, leading to the generalized $k$-means in GLMs. Our method wants to make models within clusters mostly homogeneous and models between clusters mostly heterogeneous. Because of the differences of population sizes, transportation manners, and social activities, the number of daily new cases among the fifty states and Washington DC cannot be identical or even similar, implying that it is inappropriate to compare all of the coefficients. We compare some of the coefficients. This is called the unsaturated clustering problem. In particular, we partition the coefficient vector into two sub-vectors. The first sub-vector does not contain any time information. Therefore, we only need to study the second sub-vector. We implement the generalized $k$-means to the second sub-vector only. This problem can be partially reflected by Figure~\ref{fig:six regression lines}. Suppose that six regression lines are compared. Note that the intercepts do not contain any time information. We allow them to vary within clusters. We restrict the generalized $k$-means on the slopes only, leading to two clusters.

We investigate our method with either a known or an unknown $k$. In the case when $k$ is unknown, we propose GIC to select the best $k$. We specify it to BIC and AIC. We find that BIC is more reliable than AIC in selecting the number of clusters. Therefore, we recommend using the BIC selector. We compare our method with a previous method based on the EM algorithm~\cite{qin2006}. Our simulation results show that the number of clusters can be identified by our proposed method but not by the previous method. The previous EM algorithm cannot identify the number of clusters if the true number of clusters is greater than two. We implement our method to the state-level time series data for the outbreak of COVID-19 in the United States. We find six clusters. 

The article is organized as follows. In Section~\ref{sec:method}, we propose our method. In Section~\ref{sec:asymptotic properties}, we study theoretical properties of our method. In Section~\ref{sec:simulation}, we evaluate our method with the comparison to a previous method by simulation studies. In Section~\ref{sec:application}, we implement our method to the state-level COVID-19 data in the United States. In Section~\ref{sec:discussion}, we provide a discussion.

\section{Method}
\label{sec:method}

We propose our method under a given $k$ in Section~\ref{subsec:generalized k-means in GLMs}. It can be combined with GIC~\cite{zhangli2010} to select the best $k$ when $k$ is unknown. This is introduced in Section~\ref{subsec:generalized information criterion}. In Section~\ref{subsec:specification}, we specify our method to regression models for normal data and loglinear models for Poisson data. The two models will be used in our simulation studies in Section~\ref{sec:simulation}. The loglinear model for Poisson data can be extended to a model with overdispersion for quasi-Poisson data. It will be used in clustering for the state-level COVID-19 data in the United States.  
 
 
\subsection{Generalized $k$-Means in GLMs} 
\label{subsec:generalized k-means in GLMs}

Clustering is the problem of partitioning a set of $N$ objects, denoted by ${\cal S}=\{z_1,\dots,z_N\}$, into several non-empty subsets or clusters, such that the objects within clusters are mostly homogeneous and the objects between clusters are mostly heterogeneous. In the case when  the objects can be expressed by points in an Euclidean space, the $k$-means partitions ${\cal S}$ into $k$ distinct clusters denoted by ${\cal C}=\{C_1,\dots,C_k\}$ with ${\cal C}$ given by
\begin{equation}
\label{eq:k-means}
{\cal C}=\mathop{\arg\!\min}_{\cal C}\sum_{s=1}^k\sum_{i\in C_s} \|z_i-c_s\|^2,
\end{equation}
where $c_s$ is the center point of $C_s$. The right-hand side of~\eqref{eq:k-means} is called the SSQ criterion in the $k$-means. The generalized $k$ means is proposed if the SSQ criterion is extended to an arbitrary dissimilarity measure. In particular, let $d(z,C)$ be a selected dissimilarity measure with $z$ representing an object and $C$ representing a cluster. The generalized $k$-means solves ${\cal C}$ by 
\begin{equation}
\label{eq:generalized k-means}
{\cal C}=\mathop{\arg\!\min}_{\cal C}\sum_{s=1}^k\sum_{i\in C_s} d(z_i,C_s).
\end{equation}
When $z_i$ are points in an Euclidean space, the generalized $k$-means can be the $k$-means if one chooses $d(z_i,C_s)=\|z_i-c_s\|^2$. It can also be the $k$-medians if one chooses $d(z_i,C_s)=\|z_i-c_s\|_1$.

Because $d(z,C)$ is flexible in~\eqref{eq:generalized k-means}, the generalized $k$-means can be combined with any statistical models. Suppose that $z_i$ is composed by a vector of response and a design matrix of explanatory variables. Then, we can use a statistical model to describe the relationship between the response and the explanatory variables. The response and explanatory variables can be general, implying that the generalized $k$-means can be implemented to various kinds of data, such as text, DNA strains, or images. Here, we restrict our attention to GLMs for continuous or count responses. Our task is to group the GLMs for the objects into a number of clusters. 

Suppose that $z_i$ contains a response vector ${\bm y}_i=(y_{i1},\dots,y_{in_i})^\top$ and a design matrix ${\bf X}=({\bm x}_{i1}^\top,\dots,{\bm x}_{in_i}^\top)^\top$, such that the sample size of the data is $n=\sum_{i=1}^N n_i$. Assume that $y_{i1},\dots,y_{in_i}$ are independently obtained from an exponential family distribution with the probability mass function (PMF) or the probability density function (PDF) as
\begin{equation}
\label{eq:exponential family distribution}
f(y_{ij})=\exp\left[{y_{ij}\theta_{ij}-b(\theta_{ij})\over a(\phi)}+c(y_{ij},\phi)\right],
\end{equation}
where $\theta_{ij}$ is a canonical parameter representing the location and $\phi$ is a dispersion parameter representing the scale. The linear component $\eta_{ij}$ is related to explanatory variables by $\eta_{ij}={\bm x}_{ij}^\top{\bm\beta}_i$. The link function $g(\cdot)$ connects $\mu_{ij}={\rm E}(y_{ij})=b'(\theta_{ij})$ and $\eta_{ij}$ through 
\begin{equation}
\label{eq:link function}
\eta_{ij}=g(\mu_{ij})=g[b'(\theta_{ij})]={\bm x}_{ij}^\top{\bm\beta}_i
\end{equation} 
for all $i\in\{1,\dots,N\}$ and $j\in\{1,\dots,n_i\}$, where $\theta_{ij}=h({\bm x}_{ij}^\top{\bm\beta}_i)$ is the inverse function obtained by~\eqref{eq:link function}. The variance of the response is ${\rm V}(y_{ij})=a(\phi)v(\mu_{ij})$, where $v(\mu)=b''\{h^{-1}[g(\mu)]\}$ is the variance function of the model. If the canonical link is used, then~\eqref{eq:link function} becomes $\eta_{ij}=\theta_{ij}=g(\mu_{ij})={\bm x}_{ij}^\top\beta_i$, implying that $h(\cdot)$ is the identity function. 

The MLEs of ${\bm\beta}_i$, denoted by $\hat{\bm\beta}_i$, can only be solved numerically if the distribution is not normal. A popular and well known algorithm is the iteratively reweighted least squares (IRWLS)~\cite{green1984}. The IRWLS is equivalent to the Fisher scoring algorithm. It is identical to the Newton-Raphson algorithm under the canonical link. After $\hat{\bm\beta}_i$ is derived, a straightforward method is to estimate $\phi$ by moment estimation~\cite{mccullagh1983} as 
\begin{equation}
\label{eq:moment estimation for dispersion}
a(\hat\phi)={1\over df}\sum_{i=1}^N\sum_{j=1}^{n_i} {(y_{ij}-\hat\mu_{ij})^2 \over b''[h({\bm x}_{ij}^\top\hat{\bm\beta}_i)]},
\end{equation}
where $\hat\mu_{ij}=b'[h({\bm x}_{ij}^\top\hat{\bm\beta}_i)]$ and $df$ is the residual degrees of freedom in the derivation of $\hat{\bm\beta}_i$. If $\phi$ is not present in~\eqref{eq:exponential family distribution}, then~\eqref{eq:moment estimation for dispersion} is not needed. This occurs in Bernoulli, binomial, and Poisson models. The IRWLS is the standard algorithm for fitting GLMs. It is adopted by many software packages, such as $\textsf{R}$, $\textsf{SAS}$, and $\textsf{Python}$. 

Our interest is to determine whether ${\bm\beta}_i$ can be grouped into a few clusters, such that there is ${\bm\beta}_{i}={\bm\beta}_{i'}$ if objects $i$ and $i'$ are in the same cluster or ${\bm\beta}_{i}\not={\bm\beta}_{'}$ otherwise. The regression version of this problem has been previously investigated in gene expressions by an EM algorithm~\cite{qin2006}. It wants to know whether the entire coefficient vectors can be partitioned into a few clusters. In our method, we allow a few components of ${\bm\beta}_i$ to be different within clusters, such that we only need to partition the objects based on the rest components. 

Suppose that~\eqref{eq:link function} is expressed as
\begin{equation}
\label{eq:link function two parts}
\eta_{ij}={\bm x}_{ij1}^\top{\bm\beta}_{i1}+{\bm x}_{ij2}^\top{\bm\beta}_{i2},
\end{equation}
where ${\bm x}_{ij}=({\bm x}_{ij1}^\top,{\bm x}_{ij2}^\top)^\top$ and ${\bm\beta}_{i}=({\bm\beta}_{i1}^\top,{\bm\beta}_{i2}^\top)^\top$. We want to know whether ${\bm\beta}_{i2}$ can be grouped into a few clusters. It means that we want ${\bm\beta}_{i2}={\bm\beta}_{i'2}$ if objects $i$ and $i'$ are in the same cluster or ${\bm\beta}_{i2}\not={\bm\beta}_{i'2}$ otherwise. Based on a given ${\cal C}$, the clustering model in our method is
\begin{equation}
\label{eq:model based on a given clustering}
g(\mu_{ij})={\bm x}_{ij1}^\top{\bm\beta}_{i1}+{\bm x}_{ij2}^\top{\bm\beta}_{s2}
\end{equation}
for $z_i\in C_s$. We call~\eqref{eq:model based on a given clustering} the {\it unsaturated clustering problem}. If ${\bm\beta}_{i1}$ is absent, it becomes the {\it saturated clustering problem}, the problem studied by~\cite{qin2006}. As the choice of ${\bm x}_{ij1}$ and ${\bm x}_{ij2}$ is flexible in~\eqref{eq:model based on a given clustering}, our method can be used to group GLMs based on any sub-vectors of ${\bm\beta}_i$. We use either a likelihood ratio statistic or an $F$-statistic to construct $d(z_i,C_s)$ in~\eqref{eq:generalized k-means}. 

Our method starts with a set of cluster candidates ${\cal C}$, which can be any partition of ${\cal S}$. For any given $z_i\in{\cal S}$ and $C_s\in{\cal C}$,  if $z_i\not\in C_s$, we directly use $z_i$ and $C_s$ in the derivation of $d(z_i,C_s)$; otherwise, we remove $z_i$ from $C_s$ in the derivation. Then, the resulting $C_s$ does not contain $z_i$. We obtain the likelihood ratio statistic or the $F$-statistic for
\begin{equation}
\label{eq:hypothesis testing generalized k means}
H_0: {\bm\beta}_{i2}={\bm\beta}_{s2}\leftrightarrow H_1: {\bm\beta}_{i2}\not={\bm\beta}_{s2}.
\end{equation}
The likelihood ratio statistic is used if $\phi$ is absent in~\eqref{eq:exponential family distribution}. This appears in Poisson and binomial models. The $F$-statistic is used if $\phi$ is present. This appears in normal models. For each $z_i\in{\cal S}$ and $C_s\in{\cal C}$, we calculate the $p$-value of the likelihood ratio statistic or the $F$-statistic. We assign $z_i$ to the cluster candidate with the largest $p$-value. Then, we obtain the updated set of cluster candidates, which is used in the next iteration. To ensure $C_s$ non-empty, we do not move the object with the largest $p$-value in $C_s$ to any other cluster candidates. We have the following algorithm.

\begin{algorithm}[H]
\caption{\label{alg:generalized k-means GLMs}Generalized $k$-means in GLMs}
\begin{algorithmic}[1]
\Statex{{\bf Input}: ${\cal S}=\{z_1,\dots,z_N\}$ with $z_i=\{{\bm y}_i,{\bf X}_i\}$ }
\Statex{{\bf Output}: ${\cal C}=\{C_1,\dots,C_k\}$ and the value of the likelihood ratio or the F-statistic based on the resulting ${\cal C}$}
\State Initialization: find distinct $z_{i_1},\dots,z_{i_k}$ such that they are most dissimilar, and use those to generate the initial ${\cal C}$.
\Procedure{Update Iteratively}{}
\State For each $C_s$, compute the $p$-value of $z_i$ under~\eqref{eq:hypothesis testing generalized k means} for every $z_i\in C_s$. The object with the largest the $p$-value will be remained in $C_s$.
\State For every other $z_i$ that will not be remained, compute its $p$-values under~\eqref{eq:hypothesis testing generalized k means} for every $C_s\in{\cal C}$. Assign $z_i$ to the cluster candidate with the largest $p$ value.
\EndProcedure
\State Output.
\end{algorithmic}
\end{algorithm}

Algorithm~\ref{alg:generalized k-means GLMs} has two major stages. The second stage is given by Step 2 to Step 5, which is common in many $k$-means algorithms. Therefore, we only discuss the first stage, which is given by Step 1. The goal of the first stage is to find the best initial ${\cal C}$. We do not use the usual approach adopted by many $k$-means algorithms, as they select the initial set of cluster candidates randomly. Instead, we want to make the initial ${\cal C}$ as heterogeneous as possible. At the beginning, we randomly choose the first $z_{i}$ from ${\cal S}$. We denote it as $z_{i_1}$. It is the seed for $C_1$. We calculate the $p$-value of the likelihood ratio statistic or the $F$-statistic for
\begin{equation}
\label{eq:test seeds for the second}
H_0: {\bm\beta}_{i}={\bm\beta}_{i_12}\leftrightarrow H_1: {\bm\beta}_{i2}\not={\bm\beta}_{i_12}
\end{equation}
for every $i\not=i_1$. The object $z_{i}$ with the lowest $p$-value is chosen the seed for $C_2$. It is denoted by $z_{i_2}$. Then, we incorporate the minimax principle to select the rest seeds. Suppose that $z_{i_1}$ and $z_{i_2}$ are selected. For each $i\not=i_1,i_2$, we calculate the $p$-values of the likelihood ratio statistic or the $F$-statistic based on two testing problems as
\begin{equation}
\label{eq:hypothesis testing seeds}
H_0: {\bm\beta}_{i2}={\bm\beta}_{j2}\leftrightarrow H_1:{\bm\beta}_{i2}\not={\bm\beta}_{j2},
\end{equation}
where the two testing problems are derived by taking $j=i_1$ and $j=i_2$, respectively. We obtain two $p$-values. We assign the maximum of the two $p$-values as the $p$-value of $z_i$. The object with the lowest $p$-value is chosen as the seed for $C_3$. It is denoted by $z_{i_3}$. Using this idea, we can obtain all the seeds $z_{i1},\dots,z_{ik}$. Then, we reconsider the testing problem given by~\eqref{eq:hypothesis testing seeds}. For a given $z_i$ with $i\not\in\{i_1,\dots,i_k\}$, we calculate the $p$-values of the likelihood ratio statistic or the $F$-statistic for all $j\in\{i_1,\dots,i_k\}$. We assign $z_i$ to $C_s$ if the $p$-value is maximized at $j=i_s$. After doing this for all the rest objects, we obtain the initial ${\cal C}$. Combining it with the second stage, we obtain $k$ non-empty clusters with the corresponding value and $p$-value of the likelihood ratio statistic or the $F$-statistic. 

\subsection{Generalized Information Criterion}
\label{subsec:generalized information criterion}

The generalized $k$-means introduced in Section~\ref{subsec:generalized k-means in GLMs} cannot be used if $k$ is unknown. As the likelihood function is provided in Algorithm~\ref{alg:generalized k-means GLMs}, we can used it to define a penalized likelihood function. It is used to determine the number of clusters if $k$ is unknown. The penalized likelihood approach has been widely applied in variable selection problems. Here, we adopt the well known GIC approach~\cite{zhangli2010} to construct our objective function. The best $k$ is obtained by optimizing the objective function. 

Let $\ell({\bm\omega}_{\cal C})$ be the loglikelihood of~\eqref{eq:model based on a given clustering}, where ${\bm\omega}_{\cal C}$ represents all of the parameters involved in the model. If the dispersion parameter is not present, then ${\bm\omega}$ is composed by ${\bm\beta}_{i1}$ and ${\bm\beta}_{s2}$ for all $i\in\{1,\dots,N\}$ and $s\in\{1,\dots,k\}$ only. It is enough for us to use $\ell({\bm\omega}_{\cal C})$ to define the objective function in GIC. If the dispersion parameter is present, then we need to address the impact of estimation of $a(\phi)$, because variance can be seriously underestimated in the penalized likelihood approach under the high-dimensional setting~\cite{fanguo2012}. We decide to introduce our approach based on \eqref{eq:exponential family distribution} without $a(\phi)$. We then propose a modification when it is present. 

Assume that $a(\phi)$ does not appear in~\eqref{eq:exponential family distribution}. The GIC for~\eqref{eq:model based on a given clustering} is defined as ${\rm GIC}_{\kappa}({\cal C})=-2\ell(\hat{\bm\omega}_{\cal C})+\kappa df_{\cal C}$, where $\hat\omega_{\cal C}$ is the MLE of ${\omega}$ and $df_{\cal C}$ is the model degrees of freedom under ${\cal C}$, and $\kappa$ is a positive number that controls the properties of GIC. Let $q_1$ be the dimension of ${\bm\beta}_{i1}$ and $q_2$ be the dimension of ${\bm\beta}_{i2}$. Then, $df_{\cal C}=Nq_1+kq_2$. Note that $N$ does not vary with $k$. We define the objective function in our GIC as
\begin{equation}
\label{eq:modified generalized information criterion}
{\rm GIC}_{\kappa}({\cal C})=-2\ell(\hat{\bm\omega}_{\cal C})+\kappa kq_2.
\end{equation}
The best $k$ is solved by
\begin{equation}
\label{eq:selection of the best k}
\hat k_{\kappa}=\mathop{\arg\!\min}_{k}\{{\rm GIC}_{\kappa}(\hat{\cal C}_k)\},
\end{equation}
where $\hat{\cal C}_k$ is the best grouping based on the current $k$. The GIC given by~\eqref{eq:modified generalized information criterion} includes AIC if we choose $\kappa=2$ or BIC if we choose $\kappa=\log n$. If they are adopted, then the solutions given by~\eqref{eq:selection of the best k} are denoted by $\hat k_{AIC}$ and $\hat k_{BIC}$, respectively.

We need to estimate the dispersion parameter if it is present. Because the estimator based on the current $k$ can be seriously biased, we recommending using $k+1$ as the number of clusters in the computation of the estimate of $a(\phi)$. In particular, we calculate the best ${\cal C}$ based on the current $k$ in the generalized $k$ means. We use it to compute $\hat{\bm\beta}_{i1}$ and $\hat{\bm\beta}_{s2}$ for all $i\in\{1,\dots,N\}$ and $s\in\{1,\dots,k\}$. Then, we calculate the best ${\cal C}$ by setting the number of clusters equal to $k+1$. We use~\eqref{eq:moment estimation for dispersion} to estimate $a(\phi)$. This is analogous to the full model versus the reduced model approach in linear regression, where the variance parameter is always estimated under the full model. We treat the model with $k+1$ clusters in~\eqref{eq:model based on a given clustering} as the full model, and the model with $k$ clusters as the reduced model. We estimate $a(\phi)$ based on the full model but not the reduced model. After $a(\hat\phi)$ is derive, we put it into~\eqref{eq:modified generalized information criterion} in the computation of GIC. We then use~\eqref{eq:selection of the best k} to calculate the best $k$ with the dispersion parameter. This is important in our method for regression models.

\subsection{Specification}
\label{subsec:specification}

We specify our method to regression models for normal data and loglinear models for Poisson data. Ordinary regression models are commonly used if the interest is to discover the relationship between a single continuous response variable and a number of explanatory variables. Multivariate regression models are commonly used if at least two continuous response variables are involved. Loglinear models for Poisson data and logistic linear models for binomial data are commonly used if the response is count. They have been extended to quasi-Poisson or quasi-binomial models to incorporate overdispersion. 

In ordinary regression models,~\eqref{eq:link function two parts} becomes
\begin{equation}
\label{eq:ordinary regression model}
{\bm y}_i={\bf X}_{i1}{\bm\beta}_{i1}+{\bf X}_{i2}{\bm\beta}_{i2}+{\bm\epsilon}_i,
\end{equation}
where ${\bf X}_{i1}=({\bm x}_{i11}^\top,\dots,{\bm x}_{in_i1}^\top)^\top$, ${\bf X}_{i2}=({\bm x}_{i12}^\top,\dots,{\bm x}_{in_i2}^\top)^\top$, and ${\bm\epsilon}_i\sim {\cal N}({\bf 0},\sigma^2{\bf I}_{n_i})$. With a given ${\cal C}$, the model in our generalized $k$-means becomes
\begin{equation}
\label{eq:ordinary regression model generalized k-means}
{\bm y}_i={\bf X}_{i1}{\bm\beta}_{i1}+{\bf X}_{i2}{\bm\beta}_{s2}+{\bm\epsilon}_i
\end{equation}
for $z_i\in C_s$. We treat~\eqref{eq:ordinary regression model generalized k-means} as a special case of~\eqref{eq:ordinary regression model}. The second stage in Algorithm~\ref{alg:generalized k-means GLMs} is common. We only focus on the first stage.

We select seed $z_{i_1}$ for $C_1$ randomly. Suppose that $z_{i_1},\dots,z_{i_{\tilde k}}$ have been selected as the seeds for $C_1,\dots,C_{\tilde k}$, for any $\tilde k<k$, respectively. To determine the seed for $C_{\tilde k+1}$, we calculate the dissimilarity measure between $z_{s}$ and $z_i$ for pairs $(s,i)$ with $s\in\tilde S_{\tilde k}=\{z_{i_1},\dots,z_{i_{\tilde k}}\}$ and $i\not\in\tilde S_{\tilde k}$ based on
\begin{equation}
\label{eq:ordinary regression model interaction for two}
{\bm y}_v={\bf X}_{v1}({\bm\beta}_{s1}+\delta_v{\bm\xi}_{s1})+{\bf X}_{v2}({\bm\beta}_{s2}+\delta_v{\bm\xi}_{s2})+{\bm\epsilon}_v, 
\end{equation}
where $v=s$ or $v=i$, $\delta_v$ is the dummy variable defined as $\delta_v=0$ if $v=s$ or $\delta_v=1$ if $v=i$, and ${\bm\epsilon}_v\sim{\cal N}({\bf 0},\sigma^2{\bf I}_{n_i})$ is the error vector. We calculate the $F$-statistic for 
\begin{equation}
\label{eq:test for the difference between two}
H_0: {\bm\xi}_{i2}={\bf 0}\leftrightarrow H_1: {\bm\xi}_{i2}\not={\bf 0}.
\end{equation}
Let $p_{si}$ be the $p$-value of the $F$-statistic. We define the $p$-value of the dissimilarity between $z_i$ and $\tilde S_{\tilde k}$ as
\begin{equation}
\label{eq:p-value between one and the set}
p_i=\max_{s\in\tilde S_{\tilde k}} p_{si}.
\end{equation}
We choose $z_i$ as the seed of $C_{\tilde k+1}$ if it has the lowest $p_i$ value among all objects in $\tilde S_{\tilde k}$. Therefore, $z_{i_{\tilde k+1}}$ is given by the minimax principal as
\begin{equation}
\label{eq:minimax criterion}
i_{\tilde k+1}=\mathop{\arg\!\min}_{i}p_i=\arg\!\min_i\!\max_s p_{si}.
\end{equation}
After we obtain $\tilde S_k$, the set of all of the seeds for ${\cal C}$, we calculate the $p$-value of the $F$-statistic for~\eqref{eq:test for the difference between two} for every $s\in\tilde S_k$ and $i\not\in\tilde S_k$. We assign $z_i$ to $C_s$ if $p_{si}$ is maximized at $s$. In the end, we obtain the initial ${\cal C}$. By iterating the second stage in Algorithm~\ref{alg:generalized k-means GLMs}, we obtain $\hat{\cal C}_k$ based on a given $k$. 

Because of the presence of $\sigma^2=a(\phi)$, we follow the GIC in variable selection for regression models~\cite{zhangli2010}, and propose our GIC in the generalized $k$-means for regression models as
\begin{equation}
\label{eq:regression models GIC}
{\rm GIC}_{\kappa}({\cal C})={{\rm SSE}\over\sigma^2}+\kappa kq_2,
\end{equation}
where ${\rm SSE}$ is the sum of squares of errors given by~\eqref{eq:ordinary regression model generalized k-means}. To implement~\eqref{eq:regression models GIC}, we need to estimate $\sigma^2$. If the current $k$ is used, then the estimate of $\sigma^2$ is ${\rm SSE}$ divided by residual degrees of freedom. The first term on the right-hand side of~\eqref{eq:regression models GIC} is always equal to $n-Nq_1-kq_2$, implying that we cannot use this approach to select the best $k$. To overcome the difficulty, we recommend using $k+1$ in~\eqref{eq:ordinary regression model generalized k-means} in estimating $\sigma^2$. We denote it by $\hat\sigma_{k+1}^2$. Therefore, our GIC is
\begin{equation}
\label{eq:regression models GIC modified}
{\rm GIC}_{\kappa}({\cal C})={{\rm SSE}_k\over\hat\sigma_{k+1}^2}+\kappa kq_2,
\end{equation}
where ${\rm SSE}_k$ is the SSE with $k$ clusters in~\eqref{eq:ordinary regression model generalized k-means}. This is appropriate. If the number of true clusters is less than or equal to $k$, then slightly increasing the number of clusters by $1$ would not significantly change the estimate of $\sigma^2$, implying that the second term dominates the right-hand side of~\eqref{eq:regression models GIC modified}. Otherwise, the estimate of $\sigma^2$ would be significantly reduced, implying that the first term dominates the right-hand side of~\eqref{eq:regression models GIC modified}. Therefore, the objective function in our GIC provides a nice trade-off between the SSE and the penalty function.

For Poisson data, there is ${\rm V}(y_{ij})={\rm E}(y_{ij})=\mu_{ij}$, implying that $a(\phi)=1$. Under the framework of loglinear models,~\eqref{eq:link function two parts} becomes
\begin{equation}
\label{eq:original loglinear model}
\log(\mu_{ij})={\bm x}_{ij1}^\top{\bm\beta}_{i1}+{\bm x}_{ij2}^\top{\bm\beta}_{i2}.
\end{equation}
With a given ${\cal C}$, it reduces to
\begin{equation}
\label{eq:original loglinear model cluster}
\log(\mu_{ij})={\bm x}_{ij1}^\top{\bm\beta}_{i1}+{\bm x}_{ij2}^\top{\bm\beta}_{s2}
\end{equation}
for $i\in C_s$. Analogous to the regression models, after selecting $z_{i_1}$ randomly, we investigate 
\begin{equation}
\label{eq:pair loglinear models}
\log(\mu_{vj})=x_{vj1}^\top({\bm\beta}_{s1}+\delta_v{\bm\xi}_{i1})+x_{vj2}^\top({\bm\beta}_{s2}+\delta_v{\bm\xi}_{i2})
\end{equation}
with $v=s$ or $v=i$. We measure the dissimilarity between $z_s$ and $z_i$ by the likelihood ratio statistic for~\eqref{eq:test for the difference between two}. We derive the initial ${\cal C}$ by the same idea that we have displayed in regression models. With the second stage in Algorithm~\ref{alg:generalized k-means GLMs}, we obtain $\hat{\cal C}_k$ based on a given $k$. To determine the best $k$, we choose $-2\ell(\hat\omega_{{\cal C}_k})$ as the residual deviance of~\eqref{eq:original loglinear model cluster}. As the dispersion parameter is not present, the implementation of GIC is straightforward.

For quasi-Poisson data, there is ${\rm V}(y_{ijj})=\phi{\rm E}(y_{ij})=\phi\mu_{ij}$, implying that $a(\phi)=\phi$. We can still use~\eqref{eq:original loglinear model},~\eqref{eq:original loglinear model cluster}, and~\eqref{eq:pair loglinear models} to find the best ${\cal C}$. To determine the best $k$, we estimate $\phi$ by~\eqref{eq:moment estimation for dispersion}, which is the Pearson goodness-of-fit given by~\eqref{eq:original loglinear model cluster} divided by its residual degrees of freedom. For the same reason, we choose the number of clusters equal to $k+1$ in~\eqref{eq:original loglinear model cluster} in estimating $\phi$. We denote it by $\hat\phi_{k+1}$, leading to 
\begin{equation}
\label{eq:GIC quasi-poisson}
{\rm GIC}_{\kappa}({\cal C})={G_k^2\over\hat\phi_{k+1}}+\kappa kq_2,
\end{equation} 
where $G_k^2$ is the residual deviance (i.e., deviance goodness-of-fit) with $k$ clusters in~\eqref{eq:original loglinear model cluster}. 

\section{Asymptotic  Properties}
\label{sec:asymptotic properties}

We evaluate the asymptotic properties of our method under $n=\sum_{i=1}^N n_i\rightarrow\infty$. It is achieved by letting $n_{\min}=\min_i(n_i)\rightarrow\infty$. To simplify our notations, we assume that $n_i$ are all equal to $n_0$ and $|C_s|$ are all equal to $c$ such that we have $N=kc$ and $n=kcn_0$ in our data.  The case with distinct $n_i$ and $|C_s|$ can be proven under their minimums tend to infinity with bounded ratios between the minimums and the maximums.

The asymptotic properties are evaluated under $n_0\rightarrow\infty$ with $k,c\rightarrow\infty$. For any $i\not=i'$, let $\Lambda_{ii'}$ be the likelihood ratio statistic for 
\begin{equation}
\label{eq:test two different objects}
H_0: {\bm\beta}_{i2}={\bm\beta}_{i'2}\leftrightarrow H_1: {\bm\beta}_{i2}\not={\bm\beta}_{i'2}.
\end{equation}
As $n_0\rightarrow\infty$, $-2\log\Lambda$ is asymptotically $\chi_{q_2}^2$ distributed if $z_i$ and $z_{i'}$ are in the same cluster, or goes to $\infty$ with rate $n_0$ otherwise. Because~\eqref{eq:test two different objects} is applied to all pairs $(i,i')$ in ${\cal S}$, the  multiple testing problem must be addressed. This can be solved by the method of higher criticisms~\cite{donoho2004}. 


\begin{lem}
\label{lem:distribution under the model}
Assume that $(y_{ij},{\bf x}_{ij}^\top)^\top$ for $j\in\{1,\dots,n_0\}$ are iid copies from~\eqref{eq:model based on a given clustering} for any given $i\in{\cal S}$. If $z_i$ and $z_{i'}$ are in the same cluster, then $-2\log\Lambda_{ii'}\stackrel{L}\rightarrow\chi_{q_2}^2$. If $z_i$ and $z_{i'}$ are in different clusters, then exists a positive constant $A=A({\bm\beta}_{i},{\bm\beta}_{i'},\phi)$, such that the limiting distribution of $-2\log\Lambda-n_0A$ is non-degenerate as $n_0\rightarrow\infty$.
\end{lem}

\noindent
{\bf Proof.} The conclusion can be proven by the standard approach to the asymptotic properties of maximum likelihood and M-estimation. Please refer to Chapter 22 in \cite{ferguson1996} and Chapter 5 in~\cite{vandervaart1998}. \qed


\begin{thm}
\label{thm:main theorem asymtotics given k}
If the assumption of Lemma~\ref{lem:distribution under the model} holds, and $N=o(e^{n_0^\alpha})$ for some $\alpha\in(0,1)$ when $n_0\rightarrow\infty$, then $\hat{\cal C}_k\stackrel{P}\rightarrow {\cal C}$.  
\end{thm}

\noindent
{\bf Proof.} Note that the likelihood ratio test based on $\Lambda_{ii'}$ is applied to distinct $i,i'\in{\cal C}$. We need to evaluate the impact of the multiple testing problem. We examine  the distribution of the $\max_{i\not=i'}\lambda_{ii'}$ based on Lemma~\ref{lem:distribution under the model}. According to~\cite{donoho2004}, it is asymptotically bounded by a constant times $2\log N$ if $z_i$ and $z_{i'}$ are in same clusters or increases to $\infty$ with rate $n_0$ if $z_i$ and $z_{i'}$ are in different clusters. Thus, with probability $1$, the increasing rate of $\Lambda_{ii'}$ with $z_i$ and $z_{i'}$ in different clusters is faster than that of $\Lambda_{ii'}$ with $z_i$ and $z_{i'}$ in same clusters, implying the conclusion. \qed

\begin{thm}
\label{thm:main theorem asymptotic unknown k}
Assume that $a(\phi)$ is not present in~\eqref{eq:exponential family distribution} or $a(\phi)$ is consistently estimated by $a(\hat\phi)$ used in the construction of GIC, and the assumption of Theorem~\ref{thm:main theorem asymtotics given k} holds. If $\kappa^{-1}\log c\rightarrow 0$ as $n_0\rightarrow\infty$, then $\hat k_{\kappa}\stackrel{P}\rightarrow k$ and $\hat{\cal C}_{\hat k_{\kappa}}\stackrel{P}\rightarrow {\cal C}$.
\end{thm}

\noindent
{\bf Proof.} If $\hat k_{\kappa}<k$, then we can find at least one pair of $z_{i}$ and $z_{i'}$, such that they are not in the same cluster but they are grouped to the same cluster. By Lemma~\eqref{lem:distribution under the model}, the first term on the right-hand side of~\eqref{eq:modified generalized information criterion} goes to $\infty$ with rate $n_0$. It is faster than the rate of GIC under $\hat k_{\kappa}=k$, implying that $P(\hat k_{\kappa}<k)=0$ as $n_0\rightarrow\infty$. Therefore, we only need to study the case when $\hat k_{\kappa}\ge k$. 
Note that the loglikelihood function of~\eqref{eq:model based on a given clustering} based on a given ${\cal C}$ is equal to the sum of the loglikelihood functions obtained from each $C_s\in{\cal C}$. By Theorem~\ref{thm:main theorem asymtotics given k}, we can restrict our attention to the case when all objects in $C_s$ are in the same cluster. By~\cite{donoho2004}, with probability $1$, the loglikelihood function~\eqref{eq:model based on a given clustering} in $C_s$ is not higher than that under the true cluster plus $2\log c$. By the property of the $\chi^2$-approximation of the likelihood ratio statistic under the true ${\cal C}$, we have that with probability $1$ the first term on the right-hand side of~\eqref{eq:modified generalized information criterion} is not higher than $n_0N-(Nq_1+kq_2)+2kq_2\log c$. Combining it with the second term, we conclude that $\hat k_{\kappa}\stackrel{P}\rightarrow k$. We obtain the first conclusion. Then, we draw the second conclusion $\hat{\cal C}_{\hat k_{\kappa}}\stackrel{P}\rightarrow {\cal C}$ by Theorem~\ref{thm:main theorem asymtotics given k}. \qed

Theorem~\ref{thm:main theorem asymtotics given k} implies that both $c$ and $k$ can increase exponential fast than $n_0$ if $k$ is known, but the rate is significantly reduced if $k$ is unknown. If $c\rightarrow\infty$, then we cannot choose $\kappa=2$ in our method, implying that $\hat k_{AIC}$ is not consistent, but we can still show that $\hat k_{BIC}$ is consistent. 

\begin{cor}
\label{cor:BIC consistency}
Suppose that all of assumptions of Theorem~\ref{thm:main theorem asymptotic unknown k} are satisfied. If $k\rightarrow\infty$ or $k$ is constant, and $c/n_0\rightarrow 0$ when $n_0\rightarrow\infty$, then $\hat k_{BIC}\stackrel{P}\rightarrow{k}$. 
\end{cor}

\noindent
{\bf Proof.} Note that the increasing rate of $\log n$ cannot be lower than the increasing rate of $\log c$. We draw the conclusion by Theorem~\ref{thm:main theorem asymptotic unknown k}. \qed

\section{Simulation}
\label{sec:simulation}

We carried out simulations to evaluate our methods. For an estimated cluster assignment $\hat{\cal C}$ and the true clustering assignment ${\cal C}$, we define the clustering error ($CE$) of $\hat{\cal C}$ as
\begin{equation}
\label{eq:clustering error given estimate}
CE_{\hat {\cal C}}={N\choose 2}^{-1}\#\{(i,i'): \hat\delta_{ii'}=\delta_{ii'},1\le i<i'\le N\}
\end{equation}
where $\hat\delta_{ii'}=1$ if $z_i$ and $z_{i'}$ belong to the same clusters in $\hat{\cal C}$, or $\hat\delta_{ii'}=0$ otherwise, and similarly for $\delta_{ii'}$ in ${\cal C}$. For estimated clustering assignments $\hat{\cal C}_1,\dots,\hat{\cal C}_R$ obtained from $R$ simulation replications, respectively, we calculate the percentage of clustering object errors ($OE$) by
\begin{equation}
\label{eq:percentage of clustering error}
OE={100\over R}\sum_{j=1}^R CE_{{\hat C}_j}.
\end{equation}
This is a commonly used criterion in the clustering literature~\cite{wang2010}. We also study  the percentage of number of clusters identified correctly ($IC$) as
\begin{equation}
\label{eq:number of clusters error}
IC={100\over R}\sum_{j=1}^R I(\hat k_j=k)
\end{equation}
where $\hat k_1,\dots,\hat k_R$ are the numbers of clusters obtained from $R$ simulation replications, respectively, and $k$ is the true number of clusters. We compare clustering methods based on $CE$ and $IC$.

\subsection{Regression Models}
\label{eq:regression models}

We established regression models with $k=2,3$ clusters. Each cluster had $c=10,20$ objects. Each object contained $n_0=50,100$ observations. We generated explanatory variables $x_{ij1}$ from ${\cal U}[18,70]$ and $x_{ij2}$ from ${\cal N}(0,9)$ independently. For each selected $k$, $c$, and $n_0$, we generated the normal response from 
\begin{equation}
\label{eq:regression models in simulation}
y_{ij}=\beta_{i0}+x_{ij1}\beta_{s1}+x_{ij2}\beta_{s2}+\epsilon_{ij},
\end{equation}
for $j=1,\dots,n_0$ and $i=1,\dots,N$, where $\epsilon_{ij}\sim^{iid}{\cal N}(0,\sigma^2)$ with $\sigma=0.5,1.0$ was the random error. We evaluated our methods based on AIC and BIC with the comparison to the previous EM algorithm method proposed by~\cite{qin2006}. To implement the EM algorithm, we only considered the saturated clustering problem, where we set $\beta_{i0}$ not varied within clusters. We made $\beta_{i0}=\beta_{i'0}$ if $z_i$ and $z_{i'}$ were in the same cluster in~\eqref{eq:regression models in simulation}. If $k=2$, we chose $\beta_{i0}=1$, $\beta_{i1}=-0.06$, and $\beta_{i2}=-0.01$ if $z_i$ was in the first cluster or $\beta_{i0}=1$, $\beta_{i1}=0.06$, and $\beta_{i2}=0.01$ if $z_i$ was in the  second cluster. If $k=3$, we added one more cluster with $\beta_{i0}=1$, $\beta_{i1}=-0.02$, and $\beta_{i2}=0.01$ if $z_i$ was in the third cluster. Then, we could generate data from~\eqref{eq:regression models in simulation} with either $2$ or $3$ clusters.

\begin{table}
\caption{\label{tab:percentage of correctly identified number of clusters}Percentage of number clusters identified correctly ($IC$) in regression models based on $1000$ simulation replications  with data generated from~\eqref{eq:regression models in simulation}.}
\begin{center}
\begin{tabular}{ccccccccc}\hline
  &  & &\multicolumn{3}{c}{$n=50$} &  \multicolumn{3}{c}{ $n=100$} \\\cline{4-9}
$\sigma$ & $c$ &$k$ & EM & AIC & BIC & EM & AIC & BIC \\\hline
$0.5$&$10$&$2$&$81.1$&$15.4$&${\bf 97.6}$&$72.4$&$21.4$&${\bf 98.1}$\\
&&$3$&$0.0$&$5.6$&${\bf 97.8}$&$0.0$&$6.2$&${\bf 98.9}$\\
&$20$&$2$&$75.1$&$0.5$&${\bf 88.6}$&$75.2$&$0.3$&${\bf 93.2}$\\
&&$3$&$0.0$&$0.0$&${\bf 95.9}$&$0.0$&$0.0$&${\bf 97.3}$\\
$1.0$&$10$&$2$&$75.1$&$17.3$&${\bf 96.8}$&$71.6$&$15.9$&${\bf 98.6}$\\
&&$3$&$0.0$&$7.3$&${\bf 95.7}$&$0.0$&$4.9$&${\bf 98.4}$\\
&$20$&$2$&$74.2$&$0.5$&${\bf 93.0}$&$72.2$&$0.3$&${\bf 93.8}$\\
&&$3$&$0.0$&$0.3$&${\bf 87.6}$&$0.0$&$0.0$&${\bf 96.5}$\\\hline
\end{tabular}
\end{center}
\end{table}

Table~\ref{tab:percentage of correctly identified number of clusters} displays the simulation results for the percentage of clustering number errors with respect to the previous EM algorithm,  and our AIC and BIC selectors. In all of the simulations that we ran, we found that the number of clusters reported by the EM algorithm was either $1$ or $2$, implying that it could not find the correct number of clusters if $k>2$. The true $k$ could be detected by our BIC not our AIC. 

\begin{table}
\caption{\label{tab:percentage of error regression}Percentage of clustering object errors ($OE$) in regression models based on $1000$ simulation replications with data generated from~\eqref{eq:regression models in simulation}.}
\begin{center}
\begin{tabular}{cccccccccc}\hline
  &  & \multicolumn{4}{c}{ $n=50$} &  \multicolumn{4}{c}{$n=100$} \\\cline{3-10}
  &  & \multicolumn{2}{c}{$k=2$} &\multicolumn{2}{c}{$k=3$} &\multicolumn{2}{c}{$k=2$} &\multicolumn{2}{c}{$k=3$} \\\cline{3-10}
$\sigma$ & $c$ & EM & BIC & EM &  BIC & EM & BIC & EM &  BIC\\\hline
$0.5$&$10$&$10.0$&${\bf 0.3}$&$33.6$&${\bf 0.0}$&$14.5$&${\bf 0.2}$&$33.1$&${\bf 0.3}$\\
&$20$&$12.8$&${\bf 1.3}$&$33.1$&${\bf 0.3}$&$12.8$&${\bf 0.8}$&$31.8$&${\bf 0.2}$\\
$1.0$&$10$&$13.1$&${\bf 0.4}$&$33.6$&${\bf 3.1}$&$14.9$&${\bf 0.2}$&$36.4$&${\bf 0.2}$\\
&$20$&$13.3$&${\bf 0.8}$&$34.5$&${\bf 3.8}$&$14.3$&${\bf 0.7}$&$36.5$&${\bf 0.4}$\\\hline
\end{tabular}
\end{center}
\end{table}

Table~\ref{tab:percentage of error regression} displays the simulation results for the percentage of clustering object errors based on the previous EM algorithm and our BIC selector. We did not include AIC in the table because it could not detect the correct $k$. Our result shows that BIC in the generalized $k$-means was always better that in the previous EM algorithm. Our BIC was able to find the true number of clusters with lower clustering object errors. The previous EM algorithm cannot be used to study the unsaturated clustering problem. This is an advantage of our generalized $k$-means.

\subsection{Loglinear Models}
\label{eq:loglinear models}

Similar to the regression models, we also chose $k=2,3$ clusters in loglinear models for Poisson data. Each cluster had $c=10,20$ objects. Each object contained $n_0=50,100$ observations.  We generated explanatory variables $x_{ij1}$ and $x_{ij2}$ from ${\cal N}(0,4)$ independently. For each selected $k$, $c$, and $n_0$, we independently generated the response $y_{ij}$ from ${\cal P}(\lambda_{ij})$ with 
\begin{equation}
\label{eq:poisson random variables}
\log\lambda_{ij}=\beta_{i0}+x_{ij1}\beta_{s1}+x_{ij2}\beta_{s2}
\end{equation}
for $j=1,\dots,n_0$ and $i=1,\dots,N$. We generated $\beta_{i0}$ independently from ${\cal N}(10,1)$. We set $(\beta_{11},\beta_{12})=(1,1)$ in the first cluster and $(\beta_{21},\beta_{22})=(-1,-1)$ in the second cluster. This was used if $k=2$. If $k=3$, we chose $(\beta_{31},\beta_{32})=(1,-1)$ in the third cluster. We evaluated our method based on AIC and BIC for the unsaturated clustering problem, where we varied $\beta_{i0}$ within clusters. 

\begin{table}
\caption{\label{tab:percentage of correct k poisson}Percentage of number clusters identified correctly ($IC$) in loglinear models based on $1000$ simulation replications with data generated from~\eqref{eq:poisson random variables}.}
\begin{center}
\begin{tabular}{cccccccccc}\hline
  &  & \multicolumn{4}{c}{$n=50$} &  \multicolumn{4}{c}{$n=100$} \\\cline{3-10}
  &  & \multicolumn{2}{c}{$k=2$} &\multicolumn{2}{c}{$k=3$} &  \multicolumn{2}{c}{$k=2$} &\multicolumn{2}{c}{$k=3$} \\\cline{3-10}
$\tau$ & $c$ & AIC & BIC & AIC & BIC & AIC & BIC & AIC & BIC\\\hline
$0.5$&$10$&$1.6$&${\bf 91.7}$&$0.3$&${\bf 90.9}$&$1.6$&${\bf 94.9}$&$0.4$&${\bf 93.9}$\\
&$20$&$0.1$&${\bf 71.2}$&$0.0$&${\bf 68.7}$&$0.0$&${\bf 80.1}$&$0.0$&${\bf 77.1}$\\
$1.0$&$10$&$2.1$&${\bf 92.5}$&$1.2$&${\bf 92.1}$&$1.3$&${\bf 95.3}$&$0.6$&${\bf 95.2}$\\
&$20$&$0.0$&${\bf 78.1}$&$0.0$&${\bf 76.5}$&$0.1$&${\bf 83.4}$&$0.0$&${\bf 83.0}$\\\hline
\end{tabular}
\end{center}
\end{table}

Table~\ref{tab:percentage of correct k poisson} displays the simulation results for the percentage of clustering number errors. We also found that the true $k$ could be identified by our BIC but not by our AIC. Table~\ref{tab:percentage of error loglinear} displays the results for the percentage of clustering object errors based on BIC. It shows that the percentage of clustering object errors was still low, indicating that BIC can be used to find the correct number of cluster with the low error rate. Therefore, we recommend using BIC in our generalized $k$-means if the number of clusters is unknown.  

\begin{table}
\caption{\label{tab:percentage of error loglinear}BIC for percentage of clustering object errors ($OE$) in loglinear models based on $1000$ simulation replications with data generated from~\eqref{eq:poisson random variables}.}
\begin{center}
\begin{tabular}{cccccc}\hline
  &  & \multicolumn{2}{c}{$k$ for $n=50$} &  \multicolumn{2}{c}{$k$ for $n=100$} \\\cline{3-6}
$\tau$ & $c$ & $2$ &$3$ &$2$ &$3$ \\\hline
$0.5$&$10$&$1.3$&$0.6$&$0.8$&$0.4$\\
&$20$&$4.4$&$2.1$&$3.0$&$1.5$\\
$1.0$&$10$&$1.1$&$0.5$&$0.7$&$0.3$\\
&$20$&$3.3$&$1.5$&$2.5$&$1.1$\\\hline
\end{tabular}
\end{center}
\end{table}

\section{Application}
\label{sec:application}

\begin{figure}
\centerline{\rotatebox{270}{\psfig{figure=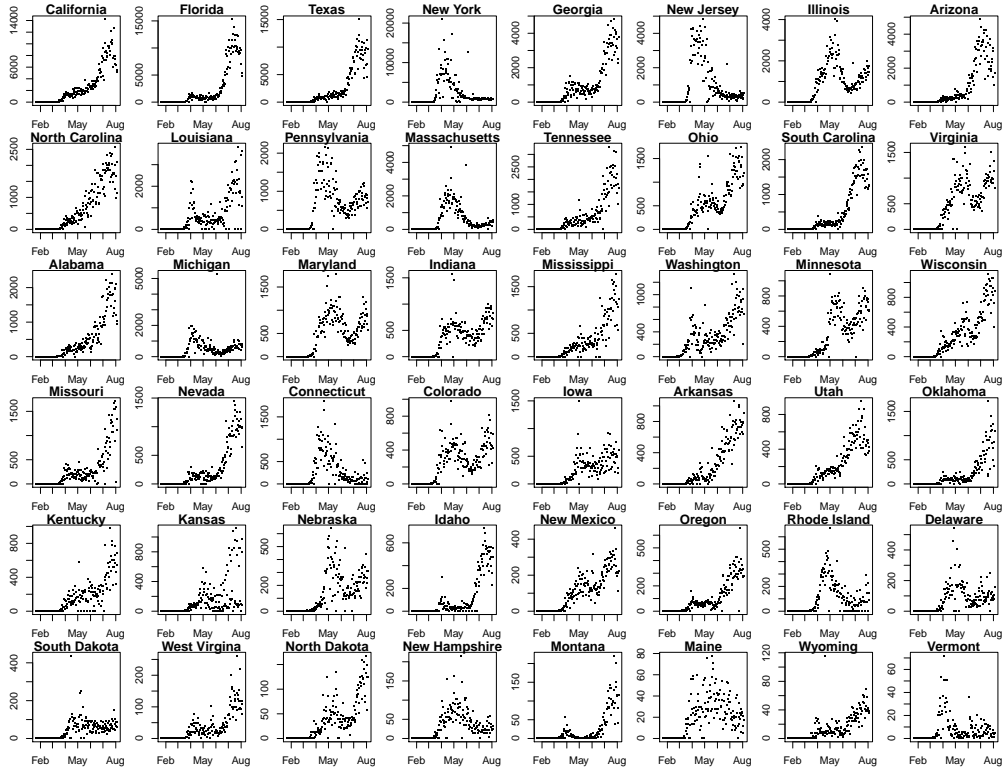,width=4.0in,}}}
\caption{\label{fig:confirmed plot}Daily new cases of COVID-19 in $48$ states in the mainland United States.}
\end{figure}

We implemented our method to the state-level COVID-19 data in the United States. It is known that the outbreak of COVID-19 in the world has occurred in March 2020 and more than $200$ countries  have affected. The situation in the United States is the most serious in the world. Until July 31, the United States has over $4.7$ million confirmed cases and one hundred sixty thousand deaths, which are the highest in the world. After briefly looking at the data (Figure~\ref{fig:confirmed plot}), we found significant changes in the time series patterns before May 31 and after June 1. We suspected two possible reasons from social medias. The first was the George Floyd issue, occurred on May 25 in Minneapolis. The second was the economy reopening issue. Most states reopened their economy or released their restrictions for the spread of the infection at the end of May. Therefore, we needed to pay attention to their impacts in the implementation of our method.

It is known that the first patient of COVID-19 appeared in Wuhan, China, on December 1 2019. In late December, a cluster of pneumonia cases of unknown cause was reported by local health authorities in Wuhan with clinical presentations greatly resembling viral pneumonia~\cite{chen2020,sun2020}. Deep sequencing analysis from lower respiratory tract samples indicated a novel coronavirus~\cite{feng2020,huang2020}. The virus of COVID-19 primarily spreads between people via respiratory droplets from breathing, coughing, and sneezing~\cite{who02272020}. This kind of spreading can cause cluster infections in society. To avoid cluster infections, many countries have imposed travel restrictions. These restrictions have affected over $91\%$ of the total population of the world with three billion people living in countries with restrictions on people arriving from other countries borders completely closed to noncitizens and nonresidents~\cite{PewResearchCenter2020}.

Exponential increasing trends are expected at the beginning of outbreaks of any infectious disease. This phenomenon has been observed in the 2009 Influenza A (H1N1) pandemic~\cite{dePicoli2011} and the 2014 Ebola outbreak in West Africa~\cite{hunt2014}. Without any prevention efforts, the exponential trend will be continuing for months until a large portion of people are infected, but this can be changed due to prevention by governments~\cite{maier2020}. 

To obtain a more appropriate model for the time series patterns in the United States, we investigate a few candidate models. We choose the response as the number of daily new cases and explanatory variables as functions of time. We find that two models were useful. The first is the exponential model as
\begin{equation}
\label{eq:exponential model}
\log\lambda_j=\mu+\beta(t_j-t_0),
\end{equation}
where $t_0$ is the starting date, $t_j$ is the current date, $\lambda_j={\rm E}(y_j)$, and $y_j$ is the number of daily new case observed on the current date. The second is the Gamma model given as
\begin{equation}
\label{eq:gamma model}
\log\lambda_j=\mu+\alpha\log(t_j-t_0)+\beta(t_j-t_0).
\end{equation}
It assumes that the expected value of number of daily new cases is proportional the density of a Gamma-distribution. If the second term is absent, then the Gamma model becomes the exponential model, implying that~\eqref{eq:exponential model} is a special case of~\eqref{eq:gamma model}.

\begin{table}
\caption{\label{tab:world table}Fitting results of the exponential and the Gamma models for the outbreak of COVID-19 in eleven selected countries between January 11 to May 31, 2020}
\begin{center}
\begin{tabular}{ccccccccc}\hline
        & \multicolumn{3}{c}{Exponential} & \multicolumn{5}{c}{Gamma} \\\cline{2-9}
Country & $\mu$ & $\beta$ & $R^2$ & $\mu$ & $\alpha$ & $\beta$ & $R^2$ & Peak \\\hline
China&$7.91$&$-0.032$&$0.368$&$-9.6$&$7.77$&$-0.290$&$0.813$&02/07\\
USA&$7.23$&$0.025$&$0.582$&$-64.7$&$20.56$&$-0.195$&$0.939$&04/26\\
Canada&$4.19$&$0.026$&$0.561$&$-75.1$&$22.61$&$-0.215$&$0.920$&04/26\\
Russia&$3.67$&$0.044$&$0.840$&$-135.2$&$37.90$&$-0.308$&$0.993$&05/13\\
Spain&$6.59$&$0.013$&$0.164$&$-77.0$&$24.72$&$-0.283$&$0.899$&04/08\\
UK&$5.53$&$0.023$&$0.446$&$-82.9$&$25.25$&$-0.248$&$0.857$&04/22\\
Italy&$6.66$&$0.010$&$0.110$&$-58.0$&$19.53$&$-0.238$&$0.945$&04/03\\
France&$6.26$&$0.012$&$0.096$&$-96.9$&$30.47$&$-0.353$&$0.694$&04/07\\
Germany&$6.33$&$0.011$&$0.103$&$-83.7$&$26.80$&$-0.317$&$0.862$&04/05\\
Switzerland&$4.90$&$0.006$&$0.030$&$-116.0$&$36.50$&$-0.463$&$0.853$&03/30\\
Sweden&$3.28$&$0.026$&$0.626$&$-43.75$&$13.6$&$-0.123$&$0.876$&04/30\\\hline
\end{tabular}
\end{center}
\end{table}

Suppose that $\alpha>0$. If $\beta>0$, then the third term dominates the variation of the right-hand side of~\eqref{eq:gamma model}. The expected value of the response goes to infinity as time goes to infinity, leading to an exponential increasing trend. If $\beta<0$, then the peak of the model is $t_{\max}=t_0-\alpha/\beta$. An increasing trend is expected if $t<t_{\max}$ and a decreasing trend is expected otherwise. Therefore, we can use the sign of $\beta$ to determine whether the outbreak is under control or out of control. 

We chose $t_0$ as January 11 in both~\eqref{eq:exponential model} and~\eqref{eq:gamma model}. We assumed that $y_i$ followed the quasi-Poisson model, such that we could fit the two models by the traditional loglinear model with the dispersion parameter $a(\phi)=\phi$ to be estimated by~\eqref{eq:moment estimation for dispersion}. We assessed the two models by their $R^2$ values, where the $R^2$ value of a GLM was defined as one minus residual deviance divided by the null deviance. We verified~\eqref{eq:exponential model} and~\eqref{eq:gamma model} by implementing them to eleven countries in the world (Table~\ref{tab:world table}), where the peak was estimated by $\hat t_{\max}=t_0-\hat\alpha/\hat\beta$ with $\hat\alpha$ and $\hat\beta$ as the MLEs of $\alpha$ and $\beta$ in the model. We found that the results given by the Gamma model were significantly better than those given by the exponential model. 

\begin{figure}
\centerline{\rotatebox{270}{\psfig{figure=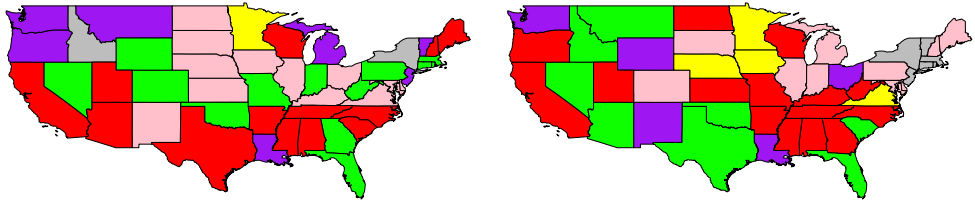,width=4.0in,}}}
\caption{\label{fig:USA clusters} Six clusters identified by BIC in generalized $k$-means for the period between and February 24 to May 31 (left) and the period between February 24 to July 31 (right), respectively.}
\end{figure}

We established our generalized $k$-means clustering under~\eqref{eq:gamma model} to group the $50$ states and Washington DC. The model was
\begin{equation}
\label{eq:gamma model generalized k-means}
\log\lambda_{ij}=\mu_i+\alpha_s\log(t_j-t_0)+\beta_s(t_j-t_0),
\end{equation}
where $\lambda_{ij}={\rm E}(y_{ij})$, $y_{ij}$ was the number of daily new cases from the $i$th state on the $j$th date, and $\alpha_s$ and $\beta_s$ were the coefficients given by the $s$th cluster. 

We looked at the state-level data and found that many of daily new cases were zeros in January and February. This was because the United States only had $6$ total number of confirmed cases until February 24. We decided to exclude data before February 24 in the analysis. We applied 
~\eqref{eq:gamma model generalized k-means} to the data between February 24 and May 31 and the data between February 24 and July 31, respectively. We compared their difference to evaluate the impact of the two issues that we mentioned at the beginning of this section. We obtained six clusters based on the BIC approach (Figure~\ref{fig:USA clusters}). 

To verify our clustering result, we examined three models. The first was the main effect model. It had only one cluster in~\eqref{eq:gamma model generalized k-means}. The second was the resulting~\eqref{eq:gamma model generalized k-means} with $k$ clusters. The third was the interaction effect model. It assumed that each state formed a cluster in~\eqref{eq:gamma model generalized k-means}. We calculated the differences of residual deviance between the first and second models, and between the first and the third models, respectively. We obtained the partial $R^2$ value by the ratio of the two differences. The partial $R^2$ value interpreted the ratio of residual deviance reduced by the model with $k$ clusters. When $k=6$, we obtained that the partial $R^2$ was $0.9235$ for data between February 24 and May 31, and $0.9606$ for data between February 24 and July 31, implying that the model with six clusters was good enough for the differences among the $50$ states and Washington DC.

\begin{table}
\scriptsize
\caption{\label{tab:fitting result gamma model}Parameter estimates in the six clusters with a selected state (State) for each cluster based on the Gamma model for the outbreak of COVID-19 in the United States, where the standard errors are given inside the parenthesis and $\times$ means out of control.}
\begin{center}
\begin{tabular}{ccccccccc}\hline
 & \multicolumn{4}{c}{02/24--05/31} & \multicolumn{4}{c}{02/24--07/31} \\\cline{2-9}
Cluster & State & $\alpha$ & $\beta$ & Peak & State & $\alpha$ & $\beta$ & Peak \\\hline
$1$ & California & $10.49(0.62)$ & $-0.8750(0.0066)$ & 5/10(2.27)  & California  & $1.958(0.25)$ & $0.0069(0.0020)$ & $\times$  \\
$2$ & New York & $24.63(0.65)$ & $-0.2962(0.0078)$ & 4/3(0.28) & New York & $11.05(0.29)$ & $-0.1206(0.0030)$ & 4/12 \\
$3$ & Illinois & $19.22(0.87)$ & $-0.1780(0.0090)$ & 4/28(0.81) & Illinois & $6.48(0.30)$ & $-0.0538(0.0026)$ & 5/11 \\
$4$ & Louisiana & $21.00(0.72)$ & $-0.2378(0.0082)$ & 4/8(0.39) & Louisiana  & $1.179(0.39)$ & $0.0044(0.0034)$ & $\times$\\
$5$ & Minnesota & $19.50(4.26)$ & $-0.1545(0.0425)$ & 5/17(7.3) & Minnesota & $8.010(0.69)$ & $-0.0548(0.0056)$ & 6/5 \\
$6$ & Florida & $19.39(0.63$ & $-0.2011(0.0068)$ & 4/26(0.42) & Florida  & $1.57(0.31)$ & $0.0178(0.0024)$ & $\times$\\\hline
\end{tabular}
\end{center}
\end{table}
\normalsize

We evaluated properties of identified clusters by the MLEs of $\alpha_s$ and $\beta_s$ with $k=6$ in~\eqref{eq:gamma model generalized k-means} with $k=6$ (Table~\ref{tab:fitting result gamma model}). We found the situation in the entire United States was under control before May 31 as the signs of $\hat{\bm\beta}_s$ were all negative. The situations in the states contained by the first, the fourth, and the six clusters became worse, but the situations in the states contained by the second, the third, and the fifth clusters were still under control. The change was probably caused by that a lot of people did not keep social distance or did not stay at home in June and July in the United States.

\section{Discussion}
\label{sec:discussion}

We have proposed a new clustering method under the framework of the generalized $k$-means to group statistical models. The method can automatically select the number of clusters if it is combined with the GIC approach. We study BIC and AIC, which are two popular special cases in GIC. Our theoretical and simulation results show that the correct number of clusters can be identified by BIC but not by AIC. Therefore, we recommend using BIC to find the number of clusters if $k$ is unknown. We implement our method to partition loglinear models for the state-level COVID-19 data in the United States and finally we have identified six clusters. An important advantage is that our method can be used to study the unsaturated clustering problem, which is different from the saturated clustering problem studied by traditional $k$-means or $k$-medians. As the choice of the dissimilarity measure is flexible, our method can be extended to many scenarios beyond GLMs. This is left to future research.

\end{document}